# CRITICAL TEST OF THE SELF-SIMILAR COSMOLOGICAL PARADIGM: ANOMALOUSLY FEW PLANETS ORBITING LOW-MASS RED DWARF STARS

Robert L. Oldershaw

Amherst College

Amherst, MA 01002

rloldershaw@amherst.edu

**Abstract:** The incidence of planetary systems orbiting red dwarf stars with masses $< 0.4\ M_\odot$ provides a crucial observational test for the Self-Similar Cosmological Paradigm. The discrete self-similarity of the paradigm mandates the prediction of anomalously few planets associated with these lowest mass red dwarf stars, in contrast to conventional astrophysical assumptions. Ongoing observational programs are rapidly collecting the data necessary for testing this prediction and preliminary results are highly encouraging. A definitive verdict on the prediction should be available in the near future.

1. Introduction

The past 16 years have been an exciting time for those who study planetary systems. Astronomers have gone from a sample of one system to a cornucopia of nearly 300 planetary objects, or exoplanets, and the discovery rate continues to rise.[1] The number of observed exoplanets is now large enough so that astronomers can *begin* to make progress in identifying the statistical characteristics of this interesting class of systems. Some of the earliest discoveries seemed to defy conventional expectations. For example, one of the earliest observations[2] of an exoplanetary system involved two planets orbiting a *pulsar*! Prior to that remarkable observation most astronomers would have regarded the possibility of pulsar/planet systems with a high degree of skepticism. Another unexpected result was the large number of systems with very massive planets orbiting extremely close to their parent stars.[3] Given the many surprises that exoplanet observations have served up over the past 16 years, we may safely anticipate that further surprises will be forthcoming as the full ranges of various properties of exoplanetary systems are explored and statistically analyzed.

In this paper we will be focusing on one important property of exoplanetary systems: the incidence of planets orbiting the lowest mass M dwarf stars, and we will be discussing this property within the context of a discrete fractal cosmological paradigm called the Self-Similar Cosmological Paradigm (SSCP).[4,5] Fractal cosmological models would appear to be quite worthy of serious consideration given the ubiquity of fractal phenomena in nature. Galaxy distributions, trees, solar intensity fluctuations, clouds, star clustering, base-pairing in DNA, river systems, stock exchange fluctuations, distributions of atoms and molecules in the interstellar medium, neuronal and circulatory systems, Brownian motions, and fluid turbulence are a small sampling of the natural phenomena wherein fractal or self-similar organization abounds.[6] In fact, it is difficult to identify *parts* of nature that do not involve some form of self-similar structures and/or temporal processes. Therefore, it would seem that fractal cosmological paradigms for the *whole* cosmos deserve serious consideration. Two recent examples of such efforts are the Scale Relativity theory of Nottale[7,8] and the Discrete Scale Relativity theory of the present author.[9] The latter theory applies to the cosmos when the discrete self-similarity of the SSCP is *exact* and nature manifests global discrete scale invariance. A two-part review[4] of the SSCP is available and the author's website[10] offers a comprehensive and readily accessible resource for studying the SSCP and Discrete Scale Relativity.

## 2. Summary of the SSCP

In order to explain the rationale for the critical prediction presented below, it is necessary to briefly summarize the general concepts and discrete self-similar scaling equations of the SSCP. This new cosmological paradigm draws attention to the general principle that nature is organized in a highly stratified hierarchical manner, with the Atomic, Stellar and Galactic Scales constituting the three well-defined and observationally accessible cosmological Scales within the observable universe.[4] A second foundational principle of the SSCP is that the cosmological Scales are rigorously self-similar to one another, such that for any fundamental system or phenomenon on a given Scale $\Psi$ there is a discrete self-similar analogue on any other Scale $\Psi \pm n$. Dimensional parameters of analogue systems on neighboring cosmological Scales, as well as *all dimensional constants*, obey the following set of discrete self-similar transformation equations.

$$L_\Psi = \Lambda L_{\Psi-1} \quad (1)$$

$$T_\Psi = \Lambda T_{\Psi-1} \quad (2)$$

$$M_\Psi = \Lambda^D M_{\Psi-1} \quad (3)$$

The terms L, T and M designate lengths, temporal periods and masses, respectively, for analogue systems on neighboring cosmological Scales $\Psi$ and $\Psi - 1$. The values of the self-similar scaling constants $\Lambda$ and D have been determined empirically and are $\cong 5.2 \times 10^{17}$ and $\cong 3.174$, respectively.[4] The value of $\Lambda^D$ is $= 1.70 \times 10^{56}$. The term $\Psi$ is a discrete index used for ordering the cosmological Scales such that $\Psi = \{\ldots -2, -1, 0, 1, 2, \ldots\}$, and the Stellar Scale is usually assigned $\Psi = 0$.

## 3. The Critical Prediction

Given a basic understanding of the SSCP, one can construct the following remarkably simple and definitive prediction with which to test the discrete fractal paradigm. Knowing the empirical values for the mass and radius of the ground state hydrogen atom, one can use Eqs. (1) and (3) to conclude unambiguously that red dwarf stars with masses ranging between 0.1 $M_\odot$ and 0.2 $M_\odot$, and peaking at 0.145 $M_\odot$, correspond to hydrogen atoms in low energy states.[4] Since the outer plasma shell structure of the star corresponds to the wavefunction of a low energy state bound electron, an additional planetary system would have to correspond to the wavefunction of a second electron in a particle-like Rydberg state with a large principal quantum number. Such a two-electron hydrogen system can exist temporarily as a bound hydride ion (H⁻), but it is quite rare within the overwhelming majority of potential physical environments. Therefore, *within the context of the SSCP*, stably bound planetary systems would be anomalously uncommon in the case of red dwarf stars with masses below 0.4 $M_\odot$. We identify this specific mass cutoff because the Stellar Scale analogue of helium, which *is* a stable two-electron system in its neutral states, is inferred to have a mass of 0.44 $M_\odot$ for $^3$He or 0.58 $M_\odot$ for $^4$He. Existing theories of stellar evolution and planet formation, on the other hand, do not predict any cutoff in the abundance of planetary systems orbiting stars with masses between 0.4 $M_\odot$ and the hydrogen burning limit at 0.08 $M_\odot$,[11] and any postdiction of such an anomaly would appear uncomfortably *ad hoc*. Since significant fractions of G and K dwarf stars appear to have planetary systems,[1] the predicted sharp decrease in the abundance of planetary systems for M dwarf stars with masses of < 0.4 $M_\odot$ would appear to be a unique and unexpected phenomenon by which the SSCP can be vindicated or falsified.

## 4. Preliminary Evidence

Because initial exoplanetary searches often preferentially targeted solar-like stars, the discovery of exoplanets was biased in favor those orbiting F, G and K dwarf stars. Current observational surveys now include more sensitive radial velocity techniques, microlensing observations, and transit detections, so that the initial biases in exoplanet detections are in the process of being corrected. Sufficient empirical evidence for a fair and definitive observational test of the prediction identified above should be available in the near future.

A preliminary indication that there might well be an anomalously low incidence of exoplanets orbiting the lowest mass M dwarf stars was recently submitted by Cumming *et al*.[12]

It should be fully appreciated that these authors were comparing the incidence of exoplanets orbiting *all* M dwarf stars, including M dwarf stars over their whole mass range, with the corresponding incidence for F, G and K dwarf stars. The results of their research lead them to conclude[12] that the incidence of gas giant exoplanets with periods of less than 2,000 days "is $r$ = 3-10 times smaller for M dwarfs than FGK dwarfs (Fig. 17), with a two sigma limit $r > 1.5$." The statistical confidence in the results still needs to be increased via larger sample sizes, and eventually it will be very important to more rigorously determine the incidence of exoplanets as a function of M dwarf mass, especially in the $0.1 M_\odot < M < 0.4 M_\odot$ range. However the preliminary results of Cumming *et al*,[12] along with previous observational results that first suggested the possibility of anomalously low incidence of exoplanets orbiting M dwarf stars (e.g., for the first 102 exoplanets discovered, K dwarf hosts were 8 times more common than M dwarf hosts), are highly encouraging from the point of view of the SSCP.

## 5. Conclusion

The predicted exoplanet incidence anomaly for the lowest mass subclass of M dwarf stars is a simple and direct test of the discrete fractal cosmological paradigm. The SSCP and more conventional astrophysical theory make very different predictions for the incidence of exoplanets stably orbiting M dwarf stars with masses of $\approx 0.15 M_\odot$. This critical test can be observationally decided in the near future once the statistics for M dwarf stars are improved by a factor of 2-3, allowing astrophysicists to distinguish the exoplanet incidence for the $M < 0.4 M_\odot$ subclass from that of the $M > 0.4 M_\odot$ subclass. If the predicted anomaly is not observed, then the concept of cosmological self-similarity would be placed in considerable doubt. On the other hand, if this unique prediction is vindicated, then the implications would be of considerable importance to the fields of astrophysics and cosmology.